\documentclass{sigplanconf}

\usepackage{amsmath}
\usepackage{acronym}
\usepackage{graphicx}
\usepackage{subfigure}
\usepackage{booktabs}
\usepackage{url}
\usepackage{color}

\acrodef{DES}{Discrete Event Simulation}
\acrodef{FEL}{Future Event List}
\acrodef{GVT}{Global Virtual Time}
\acrodef{LVT}{Local Virtual Time}
\acrodef{LP}{Logical Process}
\acrodef{PADS}{Parallel and Distributed Simulation}
\acrodef{PE}{Processing Element}

\newcommand{\msg}[1]{\ensuremath{\langle #1 \rangle}}

\begin{document}

\conferenceinfo{FHPC'12,} {September 15, 2012, Copenhagen, Denmark.}
\CopyrightYear{2012}
\copyrightdata{978-1-4503-1577-7/12/09}

\title{Parallel Discrete Event Simulation with Erlang\footnotemark}

\authorinfo{Luca Toscano \and Gabriele D'Angelo \and Moreno Marzolla}
           {Department of Computer Science, University of Bologna}
           {luca.toscano2@studio.unibo.it, g.dangelo@unibo.it, marzolla@cs.unibo.it}

\maketitle

\footnotetext{The publisher version of this paper is available at \url{http://dx.doi.org/10.1145/2364474.2364487}. \textbf{{\color{red}Please cite as: Luca Toscano, Gabriele D'Angelo, Moreno Marzolla. Parallel Discrete Event Simulation with Erlang. Proceedings of ACM SIGPLAN Workshop on Functional High-Performance Computing (FHPC 2012) in conjunction with ICFP 2012. ISBN: 978-1-4503-1577-7.}}}

\begin{abstract}
  Discrete Event Simulation (DES) is a widely used technique in which
  the state of the simulator is updated by events happening at
  discrete points in time (hence the name). DES is used to model and
  analyze many kinds of systems, including computer architectures,
  communication networks, street traffic, and others.  Parallel and
  Distributed Simulation (PADS) aims at improving the efficiency of
  DES by partitioning the simulation model across multiple processing
  elements, in order to enable larger and/or more detailed studies to
  be carried out. The interest on PADS is increasing since the
  widespread availability of multicore processors and affordable high
  performance computing clusters. However, designing parallel
  simulation models requires considerable expertise, the result being
  that PADS techniques are not as widespread as they could be.  In
  this paper we describe ErlangTW, a parallel simulation middleware
  based on the Time Warp synchronization protocol. ErlangTW is
  entirely written in Erlang, a concurrent, functional programming
  language specifically targeted at building distributed systems. We
  argue that writing parallel simulation models in Erlang is
  considerably easier than using conventional programming languages.
  Moreover, ErlangTW allows simulation models to be executed either on
  single-core, multicore and distributed computing architectures. We
  describe the design and prototype implementation of ErlangTW, and
  report some preliminary performance results on multicore and
  distributed architectures using the well known PHOLD benchmark.
\end{abstract}




\section{Introduction}

Simulation is a widely used modeling technique, which is applied to
study phenomena for which a closed form analytical solution is either
not known, or too difficult to obtain. There are many types of
simulation: in a \emph{continuous simulation} the system state changes
continuously with time (e.g., simulating the temperature distribution
over time inside a datacenter); in a \emph{discrete simulation} the
system state changes only at discrete points in time; finally, in a
\emph{Monte Carlo} simulation there is no explicit notion of time, as
it relies on repeated random sampling to compute some result.

\ac{DES} is of particular interest, since it has been successfully
applied to modeling and analysis of many types of systems, including
of computer system architectures, communication networks, street
traffic, and others. In a \acl{DES}, the system is described as a set
of interacting entities; the state of the simulator is updated by
simulation \emph{events}, which happen at discrete points in time.
For example, in a computer network simulation the following events may
be defined: (1) arrival of a new packet at a router; (2) the router
starts to process a packet; (3) the router finishes processing a
packet; (4) packet transmission starts; (5) a timeout occurs and a
packet is dropped; and so on.

The overall structure of a sequential event-based simulator is
relatively simple: the simulator engine maintains a list,
called~\ac{FEL}, of all pending events, sorted in non decreasing
simulation time of occurrence. The simulator executes the main
simulation loop; at each iteration, the event with lower timestamp $t$
is removed from the~\ac{FEL}, and the simulation time is advanced to
$t$. Then, the event is executed, which triggers any combination of
the following actions:

\begin{itemize}
\item The state of the simulation is updated;
\item Some events may be scheduled at some future time;
\item Some scheduled events may be removed from the~\ac{FEL};
\item Some scheduled events may be rescheduled for a different
  time.
\end{itemize}

The simulation stops when either the~\ac{FEL} is empty, or some
user-defined stopping criteria are met (e.g., some predefined maximum
simulation time is executed, or enough samples of events of interest
have been collected). The~\ac{FEL} is usually implemented as a
priority queue, although different data structures have been
considered and provide various degree of efficiency~\cite{Jon86}.

Traditional sequential~\ac{DES} techniques may become inappropriate
for analyzing large and/or detailed models, due to the large number of
events which can require considerable (wall clock) time to complete a
simulation run. The~\ac{PADS} discipline aims at taking advantage of
modern high performance computing architectures--from massively
parallel computers to multicore processors--to handle large models
efficiently~\cite{Fuj90}. The general idea of~\ac{PADS} is to
partition the simulation model into submodels, called~Logical
Processes (LPs)~\acused{LP} which can be evaluated concurrently by
different~\acp{PE}. More precisely, the simulation model is described
in terms of multiple interacting \emph{entities} which are assigned to
different~\acp{LP}. Each~\ac{LP} that is executed on a
different~\ac{PE}, is in practice the container of a set of
entities. The simulation evolution is obtained through the exchange of
timestamped messages (representing simulation events) between the
entities. In order to ensure that causal dependencies between events
are not violated~\cite{Lamport78}, each receiving entity must process
incoming events in non decreasing timestamp order.

We observe that multi- and many-core processor architectures are now
ubiquitous; moreover, the Cloud computing paradigm allows users to
rent high performance computing clusters using a ``pay as you go''
pricing model. The fact that high performance computing resources are
readily available should suggest that~\ac{PADS} techniques--which have
been refined to take advantage precisely of that kind of
resources--are widespread. Unfortunately, \ac{PADS} techniques have
not gained much popularity outside highly specialized user
communities.

\begin{figure*}[t]
\centering\includegraphics[width=.9\textwidth]{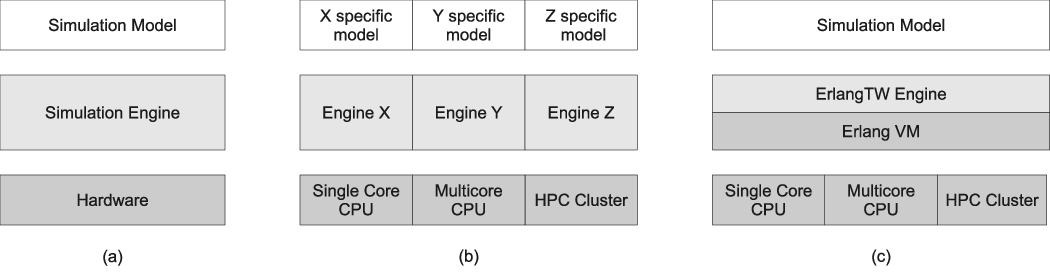}
\caption{Layered structure of discrete-event simulators}\label{fig:architecture}
\end{figure*}

There are many reasons for that~\cite{Gda11}, but we believe that the
fundamental issue with~\ac{PADS} is that parallel simulation models
are currently not transparent to the
user. Figure~\ref{fig:architecture}~(a) shows the (greatly simplified)
structure of a~\ac{DES} stack. At the higher level we have the
user-defined \emph{simulation model}; the model defines the events and
how they change the system state. In practice, the model is
implemented using either general-purpose programming language, or
languages specifically tailored for writing simulations (e.g.,
Simula~\cite{Dahl1966}, GPSS~\cite{Gordon1978},
Dynamo~\cite{Richardson1981}, Parsec~\cite{Bagrodia1998}, SIMSCRIPT
III~\cite{Rice2005}).  The simulation program depends on some
underlying \emph{simulation engine}, which provides core facilities
such as random number generation, \ac{FEL} handling, statistics
collection and so on. The simulation engine may be implemented as a
software library to be linked against the user-defined model. Finally,
at the lower level, the simulation is executed on some hardware
platform, which in general is a general-purpose processor; ad-hoc
architectures have also been considered (e.g., the ANTON
supercomputer~\cite{Shaw08}).

The current state of~\ac{PADS} is similar to
Figure~\ref{fig:architecture}~(b). Different parallel/distributed
simulation libraries and middlewares have been
proposed~(e.g.~$\mu$sik~\cite{Perumalla:2005:MPS:1069810.1070161},
SPEEDES~\cite{Steinman:2003:SPF:824475.825880}, PRIME~\cite{prime},
GAIA/ART\`IS~\cite{gda-ijspm-2009}), each one specifically tailored
for a particular environment or hardware architecture.  While hardware
dependency is unavoidable--shared memory parallel algorithms are quite
different than distributed memory ones, for example--the problem here
is that low level details are exposed to the user, which therefore
must implement the simulation model taking explicitly into account
where the model will be executed. This seriously limits the
possibility of porting the same model to different platforms.

ErlangTW is a step towards the more desirable situation shown in
Figure~\ref{fig:architecture}~(c). ErlangTW is a simulation library
written in Erlang~\cite{Armstrong07}, which implements the Time Warp
synchronization protocol for parallel and distributed
simulations~\cite{Jef85}. Erlang is a concurrent programming language
based on the functional paradigm and the actor model, where concurrent
objects interact using share nothing message passing. In this way, the
same application can potentially run indifferently on single-core
processors, shared memory multiprocessors and distributed memory
clusters. The Erlang Virtual Machine can automatically make use of all
the available cores on a multicore processor, providing a uniform
communication abstraction on shared memory machines. Also, multiple
Erlang VMs can provide a similar abstraction also on distributed
memory systems. Thanks to these features, the same ErlangTW simulation
model can be executed serially on single-core processors, or
concurrently on multicores or clusters. Of course, performance will
depend both on the model and on the underlying architecture; however,
preliminary experiments with the PHOLD benchmark (reported in
Section~\ref{sec:performance_evaluation}) show that scalability across
different processor architectures can indeed be achieved.  Moreover,
future versions of ErlangTW will add support for the adaptive runtime
migration of simulated entities (or whole~\acp{LP}) using the
serialization features offered by Erlang. An approach that, due to
many technical difficulties, is not common in~\ac{PADS} tools but that
often speeds up the simulation execution.

This paper is structured as follows. In
Section~\ref{sec:related_works} we review the scientific literature
and contrast our approach to similar works. In Section~\ref{sec:tw} we
introduce the basic concepts of distributed simulation and the Time
Warp protocol. In Section~\ref{sec:erlangtw} we present the
architecture and implementation of ErlangTW. We evaluate the
performance of ErlangTW using the PHOLD benchmark, both on a
multicore processor and on a small distributed memory cluster;
performance results are described in
Section~\ref{sec:performance_evaluation}. Finally, conclusions and
future works will be presented in Section~\ref{sec:conclusions}.

\section{Related Works}\label{sec:related_works}

Over the years, many~\ac{PADS} tools, languages and
middlewares have been proposed (a comprehensive but somewhat outdated
list can be found in~\cite{Low99}); in this section we highlight some
of the most significant results with specific attention to the 
implementations of the Time Warp synchronization mechanism.

$\mu$sik~\cite{Perumalla:2005:MPS:1069810.1070161} is a multi-platform
micro-kernel for the implementation of parallel and distributed
simulations. The micro-kernel provides advanced features such as
support for reverse computation and some kind of load balancing. 

The Synchronous Parallel Environment for Emulation and Discrete-Event
Simulation (SPEEDES)~\cite{Steinman:2003:SPF:824475.825880} and the
WarpIV Kernel~\cite{Steinman_08s-siw-025warpiv} have been used as
testbeds for investigating new approaches to parallel simulation.
SPEEDES is a software framework for building parallel simulations in
C++. SPEEDES provides support for optimistic simulations by defining
new data types for variable which can be rolled back to a previous
state (as we will see in Section~\ref{sec:tw}, this is required for
optimistic simulations). SPEEDES uses the Qheap data structure for
event management, which provides better performance with respect to
conventional priority queue data structures. SPEEDES has also been
used for many seminal works on load-balancing in optimistic
synchronization.

DSIM~\cite{chen2005} is a Time Warp simulator which targets clusters 
comprised of thousands of processors and that implements some advanced 
techniques for the memory management (e.g.~Time Quantum GVT and Local 
Fossil Collection).

We are aware of two existing simulation engines based on the Erlang
programming language: Sim94~\cite{sim94} and
Sim-Diasca~\cite{sim-diasca}. Sim94 has been originally developed for
military leadership training of battalion commanders, and is based on
a client-server paradigm. The server runs the simulation model, while
clients can connect at any time to inspect or change the simulation
state. It should be observed that Sim94 implements a conventional
sequential simulator, while ErlangTW implements a parallel and
distributed simulator based on the Time Warp synchronization protocol.
Sim-Diasca, on the other hand, is a true PADS engine (simulation
models can be executed on multiple execution units), but is based on a
time-stepped synchronization approach. A time-stepped simulation is
divided into fixed-length time steps; all execution units execute each
step concurrently and synchronize before executing the next one (see
Section~\ref{sec:tw}). Time-stepped simulations can be appropriate for
systems whose evolution is ``naturally'' driven by a sequence of steps
(e.g., circuit simulation evolving according to a global
clock). Issues in time-stepped simulations include the need to find
the appropriate duration of steps, and the high cost of
synchronization.

A recent work~\cite{gda12} investigated the use of the Go programming
language\footnote{\url{http://golang.org/}} to implement an optimistic
parallel simulator for multicore processors. The simulator, called
Go-Warp, is based on the Time Warp mechanism. Go provides mechanisms
for concurrent execution and inter-process communication, which
facilitate the development of parallel applications. Like Erlang, all
these mechanisms are part of the language core and are not provided as
external libraries. However, Go-Warp can not be executed on a
distributed memory cluster without a major redesign; with this
respect, ErlangTW represents a significant improvement, since the
simulator runs without any modification on both shared memory and
distributed memory architectures. To the best of our knowledge, Erlang
has not been used to implement a Time Warp simulation engine.

\section{Distributed Simulation}\label{sec:tw}

A~\acf{PADS} can be defined as ``a simulation in which more than one
processor is employed''~\cite{perumalla2007}. As already observed in
the introduction, there are many reasons for relying on~\ac{PADS}: to
obtain results faster, to simulate larger scenarios, to integrate
simulators that are geographically distributed, to integrate a set of
commercial off-the-shelf simulators and to compose different
simulation models in a single simulator~\cite{Fuj00}.

The main difference between sequential simulation and~\ac{PADS} is
that in the latter there is no global shared system state. A~\ac{PADS}
is realized as a set of \emph{entities}; an entity is the smallest
component of the simulation model, and therefore defines the model's
granularity. Entities interact with each other by exchanging
timestamped events. Entities are executed inside containers
called~\acp{LP}. Each~\ac{LP} dispatches the events to the contained
entities, and interacts with the other~\acp{LP} for synchronization
and data distribution. In practice, each~\ac{LP} is usually executed
by a~\ac{PE} (e.g., a single core in modern multicore processors).
Each~\ac{LP} notifies relevant events to other~\acp{LP} by sending
messages using whatever communication medium is available to
the~\acp{PE}. Each message is a pair $\msg{t, e}$, where $e$ is a
descriptor of the event to be processed, and $t$ is the simulation
time at which $e$ must be processed. Of course, the message header
includes additional information, such as the ID of the originator and
destination entities.

\begin{figure}[t]
\centering\includegraphics[width=.8\columnwidth]{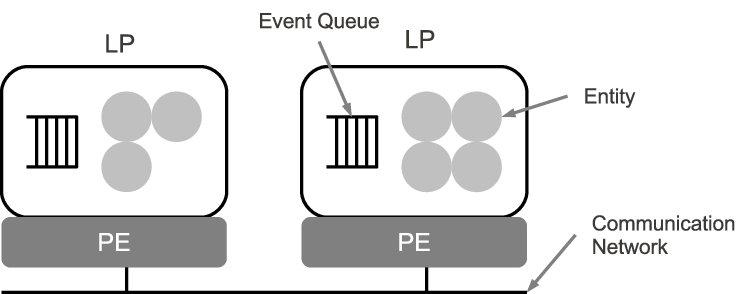}
\caption{Components of a~\ac{PADS} system}\label{fig:pads}
\end{figure}

The situation is illustrated in Figure~\ref{fig:pads}. Each~\ac{LP}
contains a set of entities, and a queue of events which are to be
executed by the local entities. The event queue plays the same role of
the~\ac{FEL} of sequential simulations: the~\ac{LP} fetches the event
with lower timestamp and forwards it to the destination entity. If an
entity creates an event for a remote entity, the~\ac{LP} uses an
underlying communication network to send the event to the
corresponding remote~\ac{LP}.

The term ``parallel simulation'' is used if the~\ac{PE} have access to
a common shared memory, or in presence of a tightly coupled
interconnection network. Conversely, ``distributed simulation'' is
used in case of loosely coupled architectures (i.e.~distributed memory
clusters)~\cite{perumalla2007}. In practice, modern high-performance
systems are often hybrid architectures where a large number of
shared memory multiprocessors are connected with a low latency
network. Therefore, the term~\ac{PADS} is used to denote both
approaches.

It is important to observe that, even if a shared system state is
indeed available on shared memory multiprocessor, the state is still
partitioned across the~\ac{PE} in order to avoid race conditions and
improve performance.

\paragraph*{Model partitioning} 
Partitioning the model is nontrivial, and in general the optimal
partition strategy may depend on the structure and semantic of the
system to be simulated. For example, in a wireless sensor network
simulation where each sensor node can interact only with neighbors, it
is reasonable to partition the model according to geographic proximity
of sensors. Many conflicting issues must be taken into account when
partitioning a simulation model into~\acp{LP}. Ideally, the partition
should minimize the amount of communication between~\acp{PE}; however,
the partition should also try to balance the workload across
different~\acp{PE}, in order to avoid bottlenecks on
overloaded~\acp{PE}. Finally, it is necessary to consider that a fixed
partitioning scheme may not be appropriate, e.g., when the
interactions among~\acp{LP} change over time. In this scenario, some
form of adaptive partitioning should be employed but this feature is not
provided by most of currently available simulators.

\begin{figure}[t]
\centering\includegraphics{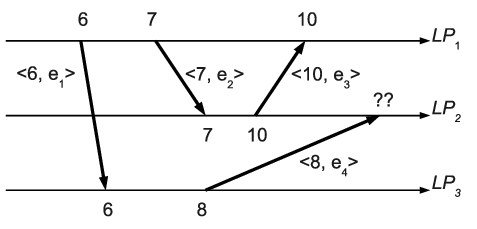}
\caption{An example of causality violation}\label{fig:causality_violation}
\end{figure}

\paragraph*{Synchronization}
The results of a~\ac{PADS} are correct if the outcome is identical to
the one produced by a sequential execution, in which all events are
processed in nondecreasing timestamp order (we assume that we can
always break ties to avoid multiple events to occur at the exact same
simulation time). In~\ac{PADS}, each~\ac{LP} $i$ keeps a local
variable $\mathit{LVT}_i$ called~\ac{LVT}, which represents the
(local) simulation time. \ac{LP} $i$ can process message $\msg{t, e}$
if $t \ge \mathit{LVT}_i$; after executing the event $e$, the~\ac{LVT}
is set to $t$.

It should be observed that the~\ac{LVT} of each~\ac{LP} advances at a
different rate, due to load unbalance or communication delays. This
may cause problems, such as the one shown in
Figure~\ref{fig:causality_violation}. We depict the timelines
associated to three~\acp{LP}, $\mathit{LP}_1$, $\mathit{LP}_2$ and
$\mathit{LP}_3$. The numbers on each timeline represents the~\ac{LVT}
of each~\ac{LP}. Arrows represent events; for simplicity, all messages
are timestamped with the sender's~\ac{LVT}.

When $\mathit{LP}_2$ receives $\msg{7, e_2}$ from $\mathit{LP}_1$, it
sets $\mathit{LVT}_2 = 7$ and executes the event $e_2$. Then,
$\mathit{LP}_2$ advances its~\ac{LVT} to 10, and sends a new message
$\msg{10, e_3}$ to $\mathit{LP}_1$. After that, message $\msg{8, e_4}$
arrives from $\mathit{LP}_3$; $e_4$ can not be executed, since
$\mathit{LVT}_2$ has already been advanced to 10. Moreover,
$\mathit{LP}_2$ sent out a message $\msg{10, e_3}$ for event $e_3$,
which may or may not have been generated should $e_4$ have been
executed before in the correct order, before $e_3$.

Figure~\ref{fig:causality_violation} shows an example of
\emph{causality violation}~\cite{Lamport78}. Two events are said to
be in causal order if one of them can have some consequences on the
other. In~\ac{PADS}, different synchronization strategies have been
developed to guarantee causal ordering of events: \emph{time-stepped},
\emph{conservative} and \emph{optimistic}.

In a \emph{time-stepped} simulation, the time is divided in fixed-size
steps, and each~\ac{LP} can proceed to the next timestep only when
all~\acp{LP} have completed the current one~\cite{1261535}.  This
approach is quite simple, but requires a barrier synchronization at
each step; the overall simulation speed is therefore always dominated
by the slowest~\ac{LP}. Furthermore, defining the ``correct'' value of
the timestep can be difficult if not impossible for some models.

The \emph{conservative} approach prevents causality violations from
occurring. A~\ac{LP} must check that no messages from the past can
arrive, before executing an event. This is achieved using the
Chandy-Misra-Bryant (CMB)~\cite{misra86} algorithm, which imposes the
following constraints: (\emph{i}) each~\ac{LP} has an incoming queue
for all other~\acp{LP} from which it can receive messages; (\emph{ii})
each~\ac{LP} must generate events in non decreasing timestamp order;
(\emph{iii}) the delivery of the events is reliable (no message can be
lost) and the network does not change the message order. Under these
assumptions, each~\ac{LP} checks all the incoming queues to determine
what is the next safe event to be processed. If there are no empty
queues, then the incoming event with lower timestamp is safe and can
be executed. Unfortunately, this mechanism is prone to deadlock, since
a~\ac{LP} can not identify the next safe event if all incoming queues
are nonempty. To avoid this, the CMB algorithm introduces a new type
of message (called NULL messages) with no semantic content. The
receipt of a NULL message $\msg{t, \mathit{NULL}}$ informs the
receiver that the sender has set its~\ac{LVT} to $t$, and hence will
not send any event with timestamp lower than $t$. NULL messages can be
used to break deadlocks, at the cost of increasing the network
load. Moreover, generation of NULL messages requires some knowledge of
the simulation model, and therefore can not be transparent the user.

Finally, the Time Warp protocol~\cite{Jef85} implements the so called
\emph{optimistic} synchronization approach.  In Time Warp,
each~\ac{LP} can process incoming events as soon as they are received.
Obviously, causality violations may happen, and special actions must
be taken to fix them. If a~\ac{LP} receives a message (called
\emph{straggler}) with timestamp smaller than some event already
processed, it must roll back the computations for these events and
re-execute them in the proper order. The problem is that some of the
events to be undone might have sent messages (events) to
other~\acp{LP} (e.g., $\msg{10, e_3}$ in
Figure~\ref{fig:causality_violation}). These messages must be
invalidated by sending corresponding \emph{anti-messages}. The
recipient of an anti-message $\msg{t, \bar{e}}$ must roll back its
state as well, which might trigger a cascade of rollbacks that brings
back the simulator to a previous state, discarding the incorrect
computations that have been performed.

In order to support rollbacks, each~\ac{LP} must keep a log of all
processed events and all messages sent, together with any information
needed to undo their effects. Obviously, logging all and every event
since the beginning of the simulation is infeasible, due to the huge
memory requirement. For this reason, the simulator periodically
computes the~\ac{GVT}, which is a lower bound on the timestamp of any
future rollback. The~\ac{GVT} is simply the smallest timestamp among
unprocessed and partially processed messages, and can be computed with
a distributed snapshot algorithm~\cite{Fuj00}. Once the~\ac{GVT} has
been computed and sent to all~\acp{LP}, logs older than~\ac{GVT} can
be reclaimed. \ac{GVT} computation can be a costly operation, since it
usually involves some form of all-to-all communications. Therefore,
finding the optimal frequency of this operation is a critical aspect
of Time Warp and typically the chosen frequency is the result of a
tradeoff between memory consumption for the logs and simulation speed.
However, when the underlying execution architecture provides efficient
support for reduction operations, the~\ac{GVT} computation does not
add too much overhead, and the Time Warp protocol can achieve almost
linear speedup even on very large setups~\cite{Perumalla11}.

Optimistic synchronization offers some advantages with respect to
conservative approaches: first, optimistic synchronization is
generally capable of exploiting a higher degree of parallelism;
second, conservative simulators require model specific information in
order to produce NULL messages, while optimistic mechanisms are less
reliant on such information (although they can exploit it if
available)~\cite{Fujimoto:1989:PDE:76738.76741}. 

\section{The ErlangTW Simulator}\label{sec:erlangtw}

Erlang is a functional, concurrent programming language based on
lightweight threads (LWT) and message passing. This makes it well
suited for developing parallel applications both on shared memory
multicore machines and on a distributed memory cluster. An Erlang
program is compiled to an intermediate representation called BEAM,
which is executed on a Virtual Machine. If Symmetric Multiprocessing
is enabled, the VM creates a separate scheduler for each CPU core;
each scheduler fetches one ready LWT from a common queue and executes
it. The \verb+spawn+ function can be used to create a new thread
executing a given function. The VM will take care of dispatching
threads to active schedulers. The fact that there is no 1:1 mapping
between LWT and OS threads facilitates the work of the developer,
since the VM takes care of balancing the load across the available
processors.

Each LWT has an identifier that is guaranteed to be unique across all
VM instances, even those running on different hosts connected through
a network. The identifier can be used by send/receive primitives,
which are provided directly by the language itself and do not require
external libraries.

The ErlangTW Simulator is an implementation of the Time Warp algorithm
described in Section~\ref{sec:tw}. Although Time Warp requires fairly
sophisticated state management capabilities to support rollbacks and
antimessages, it turned out that this (fairly limited) complexity is
paid back by the fact that Time Warp does not require ad-hoc
modifications of simulation models (e.g., to compute NULL events).

\paragraph*{Message Format}
Messages exchanged between~\acp{LP} are represented using the record
data type, providing the abstraction of a key-value tuple. Messages
have the following structure:

\begin{verbatim}
-record(message, {type, 
                  seqNumber, 
                  lpSender, 
                  lpReceiver, 
                  payload, 
                  timestamp}).
\end{verbatim}

The \verb+type+ field represents the message type; current types are:
\emph{event} (normal event), \emph{ack} (acknowledgement to ensure
reliable delivery of messages), \emph{marked\_ack} (special kind of
acknowledgement required by the Samadi's algorithm, described later),
and \emph{antimessage} (used during rollbacks).  \verb+seqNumber+ is a
numeric value representing how many messages the sender~\ac{LP} has
sent, \verb+lpReceiver+ and \verb+lpSender+ are the unique identifiers
of the sender and receiver~\ac{LP}.  \verb+payload+ is the actual
content of the message, describing the event to process and all
ancillary data. Finally, \verb+timestamp+ is simulated time associated
to the event contained in the payload.  The simulator needs to
acknowledge messages in order to guarantee the correctness of its
global state, because each message in the system must be taken into
account by one~\ac{LP} only. The Erlang VM guarantees message
delivery, but only from an LWT to another one's mailbox, therefore
this could lead to the situation in which an~\ac{LP} has received a
particular message but it has not already read it, so it is unaware of
its presence. Conversely once an~\ac{LP} receives an acknowledge for a
message it knows that it has already been taken into account by the
receiver. An example of global state is the Global Virtual Time,
explained in the following.

Here an example of a message:

\begin{verbatim}
#message{type=event/ack/marked_ack/antimessage, 
         seqNumber=100, 
         lpSender=<100,0,0>,
         lpReceiver=<100,1,0>, 
         payload="hello", 
         timestamp=10}
\end{verbatim}

\paragraph*{Event Queue}
Each~\ac{LP} maintains a priority queue of incoming messages sorted in
nondecreasing timestamp order. The~\ac{LP} fetches the message with
lower timestamp from the queue and, if the message is not a straggler,
immediately executes the associated event. The queue is implemented as
an Andersson General Balanced Tree~\cite{And99}. The tree contains
(Key, Value) pairs, where the Key is the simulation time, and the
Value is a list of events which are scheduled to happen at that time
(ErlangTW supports simultaneous events, i.e., multiple events
happening at the same simulated time).

\paragraph*{Logical Processes}
Each~\ac{LP} is implemented as an Erlang LWT created using the
\verb+spawn+\ function. \acp{LP} communicate using the \verb+send+ and
\verb+receive+ operators. The state of an~\ac{LP} is kept in a record
with the following structure:

\begin{verbatim}
-record(lp_status, {my_id, 
                    received_messages, 
                    inbox_messages, 
                    max_received_messages,
                    proc_messages,  
                    to_ack_messages, 
                    anti_messages, 
                    current_event,
                    history, gvt, 
                    rollbacks, 
                    timestamp,
                    model_state, 
                    init_model_state,
                    samadi_find_mode, 
                    samadi_marked_messages_min, 
                    messageSeqNumber, 
                    status}).
\end{verbatim}

\noindent where:
\begin{description}
\item[my\_id] is the unique identifier of the~\ac{LP};
\item[received\_messages] is the list of unprocessed messages, read from the
process mailbox;
\item[inbox\_messages] is the incoming message queue containing
  unprocessed messages;
\item[proc\_messages] is a data structure which contains, for each
  processed event, the list of messages sent by that event to remote
  entities. This data structure is required to perform rollbacks when
  necessary, because it contains the event to reprocess and the
  antimessages to send;
\item[to\_ack\_messages] is a list of events, sorted in nondecreasing
  timestamp order, related to the messages sent by the~\ac{LP} still
  to be acknowledged;
\item[model\_state] is the user-defined structure containing the state
  of the simulation model;
\item[timestamp] is the~\ac{LVT};
\item[history] is the list of processed events, used by the Time Warp
  protocol to perform rollbacks when necessary. Each element of the
  list is a tuple of the form \{Timestamp, model\_state, Event\}, and
  record the state of this~\ac{LP} at the given simulation time, before
  the Event has been processed. A tuple is added to the history after an event has been
  extracted from \emph{inbox\_messages} and executed;
\item[samadi\_*] data structures needed in order to implement the Samadi's GVT algorithm, as stated in
the next paragraph.
\end{description}

\paragraph*{Implementing Simulated Entities}
As already described in Section~\ref{sec:tw}, a~\ac{LP} is a container
of simulation entities. Each entity is the representation of some
actor or component of the ``real'' system. By decoupling~\acp{LP} from
entities, the simulation modelers can avoid dealing with partitioning;
however, if more control over the simulator is desired, the modelers
can implement their own custom partitioning by working at the~\ac{LP}
level.

In ErlangTW there is a layer between~\ac{LP} and entities, in order to
implement the separation of concerns described above. The modeler
implements three methods in a particular Erlang module called
\texttt{user}; these methods define the actions executed by
each~\ac{LP} during initialization, event processing, and
termination. The PHOLD model (described in
Section~\ref{sec:performance_evaluation}) uses an initialization
function to evenly partition the entities between the
running~\acp{LP}. The event processing function implements the
behavior executed by each entity upon receipts of a new
message. Finally, the termination function is normally used to display
or save simulation results or other information at the end of each
simulation run. Each message contains a field called \verb+payload+
that could transport any kind of user-defined data. As a specific
example, the event data structure used by the PHOLD model to manage
entities has the following structure:

\begin{verbatim}
-record(payload, 
    {entitySender, entityReceiver, value}).
\end{verbatim}

\noindent and can be instantiated, for example, as follows:

\begin{verbatim}
#payload{entitySender=10, 
         entityReceiver=122, 
         value=42}
\end{verbatim}

In this example entity 10 has sent a message to the entity 122 with a
payload containing the integer 42. In the current implementation of
ErlangTW, where the allocation of entities on~\acp{LP} must be
manually defined, the user specifies a mapping function which is used
by ErlangTW to deliver message to the appropriate~\ac{LP}. In future
versions we plan to implement some automatic allocation mechanism and
to provide this binding transparently.

\paragraph*{Global Virtual Time}
The Global Virtual Time is calculated with Samadi's
algorithm~\cite{SMP87}. One LWT, called~\ac{GVT} Controller, is
responsible to periodically checking the smallest timestamp of all
events stored in the queues of all~\acp{LP}; the~\ac{GVT} controller
is also responsible for starting and stopping the simulation. In the
current version of ErlangTW, the~\ac{GVT} controller periodically
broadcasts a GVT computation request message to all~\acp{LP};
each~\ac{LP} sends back the value of the~\ac{LVT} such that the
controller can compute the~\ac{GVT} as the minimum of these values.
The~\ac{GVT} is finally sent to all~\acp{LP}, which can then prune
their local history by removing all checkpoints older than
the~\ac{GVT}.

In practice, the calculation of the GVT is complex given that 
some messages could be in flight when the sender and/or the receiver 
LPs are reporting their~\ac{LVT}. Ignoring these messages would 
result in a wrong (overestimated)~\ac{GVT} and hang the whole simulation.
The solution proposed by Samadi is to add an acknowledgment for each 
message used for the GVT calculation, to properly identify in flight 
messages and to decide what~\ac{LP} must take them in account.

In future versions of ErlangTW we plan to compute the~\ac{GVT} using a 
more scalable reduction operation.

\paragraph*{Random Number Generation}
The pseudo random number generator used by each simulated entity is
the Linear Congruential Generator described by Park and Miller
in~\cite{PM88}.  The initial seed can be stored in a configuration
file which is read by ErlangTW before starting the simulation
run. Each entity within the same \ac{LP} shares a common random number
generator, whose seed is initialized with the seed in the
configuration file. In this way it is possible to start the simulator
in a known state, to achieve determinism and repeatability.

\section{Performance Evaluation}\label{sec:performance_evaluation}

In this section we evaluate the scalability of ErlangTW, both on
shared memory and distributed memory architectures, using a synthetic
benchmark called PHOLD~\cite{Fuj90}, which is specifically designed
for the performance evaluation of Time Warp implementations.

\paragraph{The PHOLD Benchmark}
PHOLD is the parallel version of the HOLD benchmark for event
queues~\cite{Jon86} and it is quite simple to implement and
describe. The model is made by a set of $E$ entities that are
partitioned among $L$~\acp{LP}; each~\ac{LP} contains the same number
$E/L$ of entities. Each entity produces and consumes events. When an
entity consumes an event, a new event is generated and delivered to
another entity (note that the total number of events in the system
remains constant). The timestamp of the new event is computed by
adding an exponentially distributed random number with mean 5.0 to the
timestamp of the receiving event. In this model the recipient is
randomly chosen using a uniform distribution. Therefore, each event
has a probability $1/L$ of being sent to an entity in the same~\ac{LP}
as the originator, and a probability $(L-1)/L$ of being sent to an
entity on a different~\ac{LP}. As the number $L$ of~\ac{LP} increases,
the ratio of remote vs local events increases. The PHOLD benchmark is
homogeneous in terms of load assigned to the LPs: all of them have the
same amount of communication and computation. While this can be
unrealistic for general simulation models, it is important to remark
that the Time Warp mechanism (in its original version) does require a
good level of balancing to obtain good performance
results~\cite{Carothers2000,Gda11}. Hence, the goal of PHOLD is to
study the scalability of Time Warp implementations by considering an
appropriate execution environment.

There are four main parameters which are used to control the
benchmark:
\begin{itemize}
\item The number $L$ of~\ac{LP}
\item The number $E$ of entities
\item The \emph{event density} $\rho$, $0 < \rho \leq 1$, defined as
  the fraction of entities that generate an event at the beginning of
  the simulation. At each simulation time there are $\rho E$ events in
  the system
\item The \emph{workload}, used to tune the computation /
  communication ratio by running some CPU-intensive computation each
  time an event is processed. In our case, we implemented the workload
  as a pre-defined number of floating point operations (FPops)
\end{itemize}

\begin{table}[t]
\centering%
\begin{tabular}{rl}
Number of~\acp{LP} ($L$)   & $1, \ldots, 8$ (shared memory) \\
                           & 1, 2, 3, 6 (distributed memory)\\
Number of entities ($E$)   & 840, 1680, 2520, 3360 \\
Event Density $(\rho$)     & 0.5 \\
Workload                   & 1000, 5500, 10000 FPops \\
\end{tabular}
\caption{Parameters used in the simulations}\label{tab:parameters}
\end{table}

\paragraph{Experimental Setup}
Table~\ref{tab:parameters} shows the parameters which have been used
in the simulation runs. We tested ErlangTW both on a shared memory and
on a distributed memory architecture.

The number of entities $E$ has been chosen as multiples of 840, which
is the minimum common multiple of the number of~\acp{LP} we considered
(i.e., 840 is an integer multiple of all integers in the range $1,
\ldots, 8$). This ensures that the number of entities allocated to
each~\ac{LP}, $E/L$, is an integer.

As already described, the event density has been set to 0.5, which
means that, at a given time, the average number of events in the system is $0.5 \times E$.  We
considered three different workloads of 1000, 5500 and 10000 Floating
Point Operations. Finally, the~\ac{GVT} is computed every 5 seconds.

We measured the wall clock time of a simulation run until the~\ac{GVT}
reaches 1000. In order to produce statistically valid results, we
perform 30 runs for each experiment, and compute the average of each
batch. We investigate the scalability of ErlangTW by computing the
speedup as a function of the number $L$ of~\ac{LP}.

\begin{table*}[t]
\centering\small
\begin{tabular}{rllllll}
\toprule
{\bf Host} & {\bf CPU} & {\bf Physical Cores} & {\bf HT} & {\bf RAM} & {\bf Operating System} & {\bf Network} \\
\midrule
\texttt{gda i7} & Intel i7-2600 3.40GHz & 4 & Yes & 8GB & GNU/Linux Kernel 3.2 (x86 64) & Not used \\
\midrule
\texttt{cassandra} & Intel Xeon 2.80GHz & 2 & Yes & 3GB & GNU/Linux Kernel 2.6 (x86 32) & Gigabit Ethernet \\
\texttt{cerbero} & Intel Xeon 2.80GHz & 2 & Yes & 2GB & GNU/Linux Kernel 2.6 (x86 32) & Gigabit Ethernet\\
\texttt{chernobog} & Intel Xeon 2.40GHz & 4 & No & 4GB & GNU/Linux Kernel 2.6 (x86 64) & Gigabit Ethernet\\
\bottomrule
\end{tabular}
\caption{Experimental testbeds (top: shared memory; bottom: distributed memory)}\label{tab:testbed}
\end{table*}

\begin{figure*}[ht]
\centering%
\subfigure{\includegraphics[width=.33\textwidth]{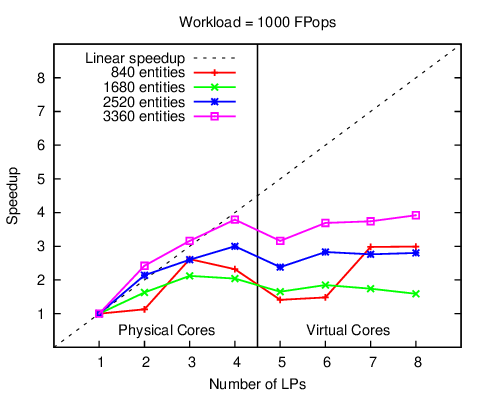}}
\subfigure{\includegraphics[width=.33\textwidth]{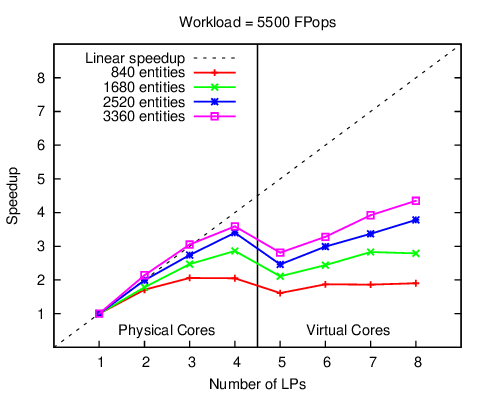}}
\subfigure{\includegraphics[width=.33\textwidth]{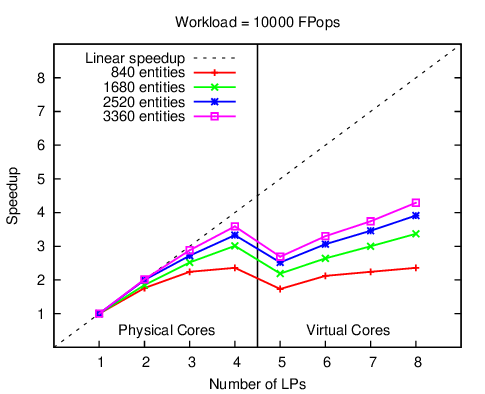}}
\caption{Speedup on the shared memory architecture as a function of
  the number of~\acp{LP} (higher is better)}\label{fig:speedup_shared}
\end{figure*}

\begin{figure*}[ht]
\centering%
\subfigure{\includegraphics[width=.33\textwidth]{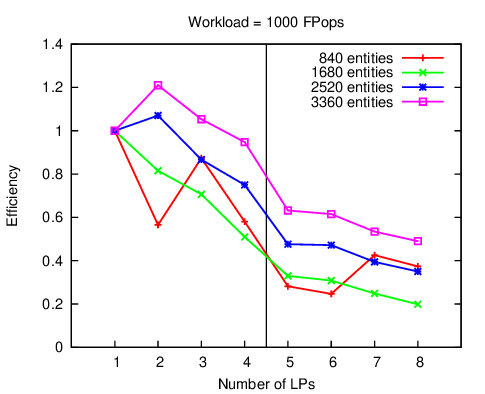}}
\subfigure{\includegraphics[width=.33\textwidth]{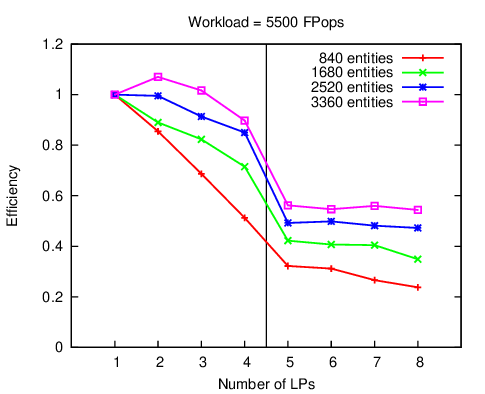}}
\subfigure{\includegraphics[width=.33\textwidth]{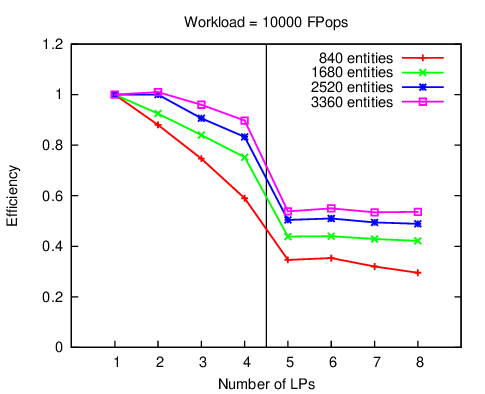}}
\caption{Efficiency on the shared memory architecture as a function of
  the number of~\acp{LP} (higher is better)}\label{fig:efficiency_shared}
\end{figure*}

\begin{figure*}[h!t]
\centering%
\subfigure{\includegraphics[width=.33\textwidth]{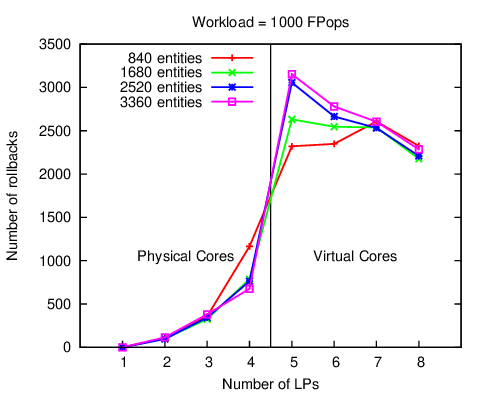}}
\subfigure{\includegraphics[width=.33\textwidth]{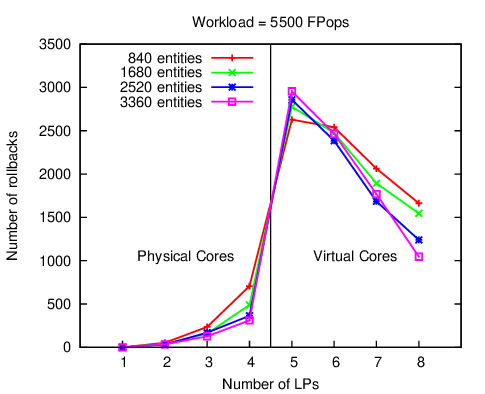}}
\subfigure{\includegraphics[width=.33\textwidth]{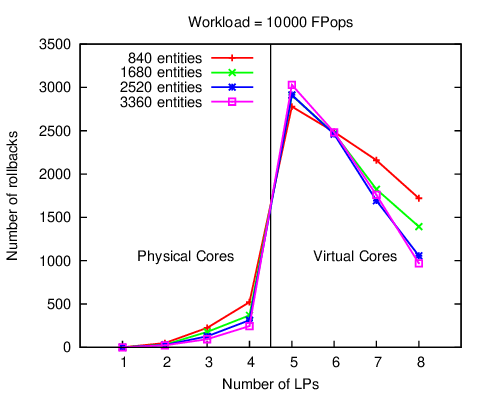}}
\caption{Total number of rollbacks on the shared memory architecture
  as a function of the number of LPs (lower is
  better)}\label{fig:rollbacks_shared}
\end{figure*}

\paragraph{ErlangTW on Shared Memory}
The shared memory system (\verb+gda i7+) is an Intel(R) Core(TM)
i7-2600 CPU \@ 3.40GHz with 4 physical cores with Hyper-Threading (HT)
technology~\cite{HT}. The system has 8 GB of RAM and runs Ubuntu 12.04
(x86\_64 GNU/Linux, 3.2.0-24-generic \#39-Ubuntu SMP). For this system
we considered several values for $L$, namely $L=1, \ldots, 8$
\acp{LP}. HT works by duplicating some parts of the processor except
the main execution units. From the point of view of the Operating
System, each physical processor core corresponds to two logical
processors. The impact of virtual cores on~\ac{PADS} is worth
investigation~\cite{gda-dsrt-2006,gda12} and will be reported in the
following.

Figure~\ref{fig:speedup_shared} shows the speedup $S_L$ as a function
of $L$; recall that $S_L = T_1 / T_L$, where $T_n$ is the wall clock
simulation time when $n$~\acp{LP} are used.  In each figure we
consider a specific value for the workload, and we plot a curve for
each number of entities $E$. As a general trend we observe that
scalability improves as the number of entities gets large; also,
scalability improves marginally if the workload (FPops) increases.
Figure~\ref{fig:efficiency_shared} shows the efficiency
$\mathit{Eff}_L = S_L / L$ as a function of $L$. The efficiency is an
estimate of the fraction of actual computation performed by all
processors, as opposed to communication and synchronization.

ErlangTW exhibits good scalability and efficiency up to $L=4$, since
in this case each~\ac{LP} can be executed on a separate physical
processor core. The transition from $L=4$ to $L=5$ shows a noticeable
drop of the speedup (and therefore in the efficiency), which is easily
explained by the effect of HT. When $L=5$, one of the physical CPU
cores executes two~\acp{LP} and becomes the bottleneck. The Time Warp
protocol works well when the workload is well balanced, but degrades
significantly if hot spots are present~\cite{Carothers2000}.

To better understand this, we report in
Figure~\ref{fig:rollbacks_shared} the mean total number of rollbacks
which occurred during the whole simulation run. A large number of
rollbacks indicates that the~\ac{LVT} at the individual~\acp{LP} are
advancing at different rates. The PHOLD model is
balanced by construction, since all entities perform identical tasks
and are uniformly distributed across the~\acp{LP}. From
Figure~\ref{fig:rollbacks_shared} we see that the number of rollbacks
increases in the region $L=1, \ldots, 4$; if $L=1$ no rollbacks
happen, since all events are managed through the event queue of a
single~\ac{LP}, so that causality is always ensured.  Adding
more~\acp{LP} increases the possibility of receiving a straggler. From
$L=4$ to $L=5$ load unbalance occurs and the number of rollbacks
sharply increases. The~\acp{LP} running on the overloaded processor
core lag behind the other~\acp{LP}, and a large number of antimessages
is produced to undo the updates performed by the faster~\acp{LP}.  As
the number of~\acp{LP} further increase, we observe that the number of
rollbacks decreases, since the system becomes more and more balanced.

In practice it is extremely difficult, if not impossible, to
statically partition a \ac{PADS} models such that the workload is
balanced across the~\ac{LP}, since the computation / communication
ratio can change during the simulation. If detailed knowledge of the
simulation model is not available in advance, as it is the case most
of the times, it is necessary to resort to adaptive entity migration
techniques to balance the~\acp{LP}~\cite{gda-ijspm-2009}. It is worth mentioning that
Erlang offers native support for code migration, which greatly
simplify the implementation of such techniques; this will be the focus
on future extensions of this work.

\begin{figure*}[ht]
\centering%
\subfigure{\includegraphics[width=.33\textwidth]{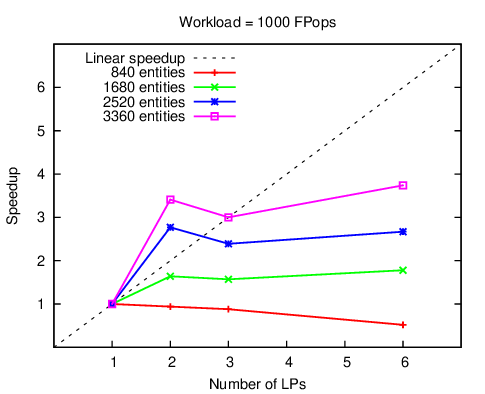}}
\subfigure{\includegraphics[width=.33\textwidth]{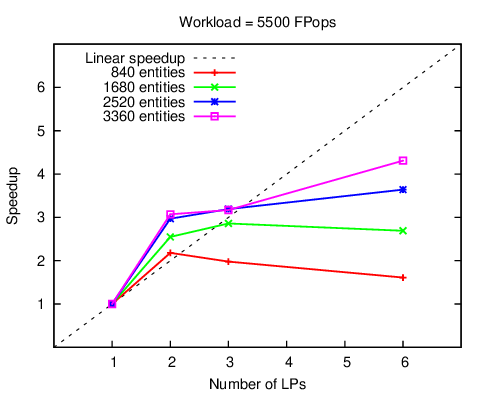}}
\subfigure{\includegraphics[width=.33\textwidth]{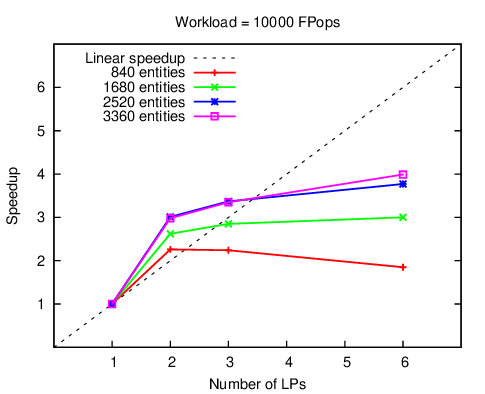}}
\caption{Speedup on the distributed memory cluster as a function of the number of LPs (higher is better); the~\ac{GVT} is computed every 5$s$ of wall clock time}\label{fig:speedup_distributed_5s}
\end{figure*}

\begin{figure*}[ht]
\centering%
\subfigure{\includegraphics[width=.33\textwidth]{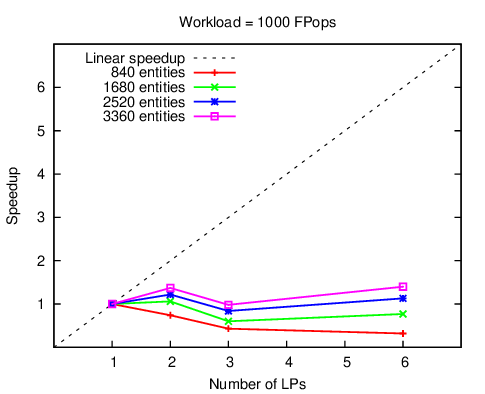}}
\subfigure{\includegraphics[width=.33\textwidth]{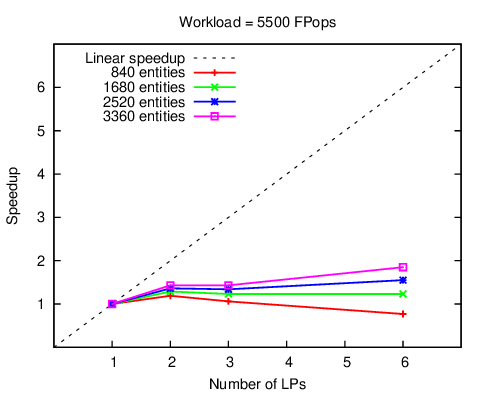}}
\subfigure{\includegraphics[width=.33\textwidth]{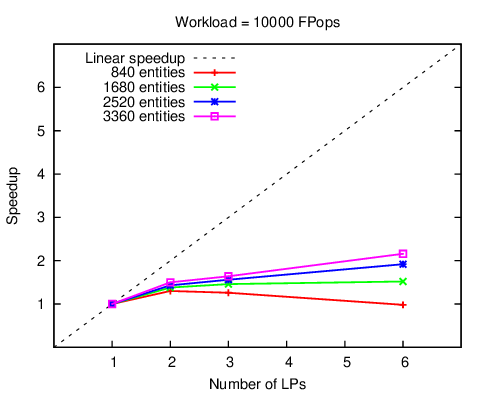}}
\caption{Speedup on the distributed memory cluster as a function of the number of LPs (higher is better); the~\ac{GVT} is computed every 1$s$ of wall clock time}\label{fig:speedup_distributed_1s}
\end{figure*}

\begin{figure*}[ht]
\centering%
\subfigure{\includegraphics[width=.33\textwidth]{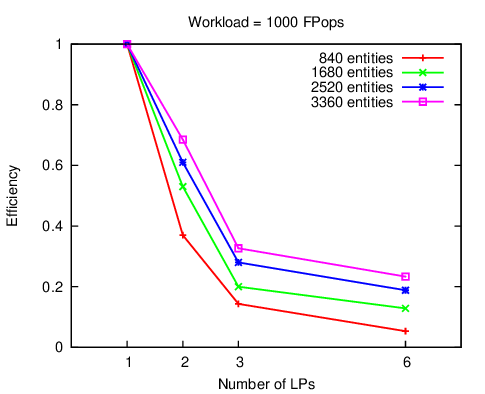}}
\subfigure{\includegraphics[width=.33\textwidth]{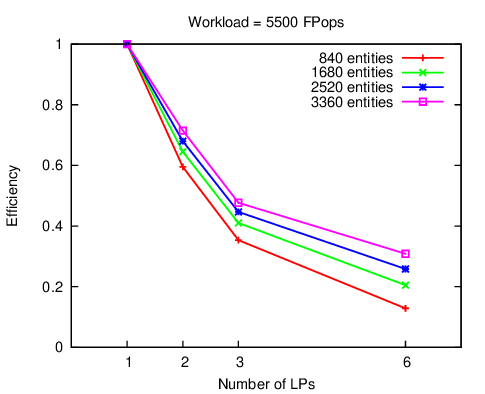}}
\subfigure{\includegraphics[width=.33\textwidth]{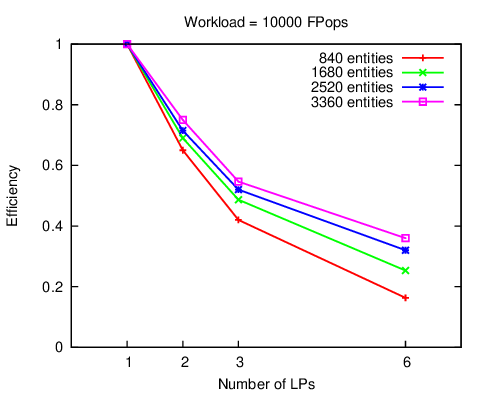}}
\caption{Efficiency on the distributed memory cluster as a function of
  the number of~\acp{LP} (higher is better)}\label{fig:efficiency_distributed_1s}
\end{figure*}

\begin{figure*}[ht]
\centering%
\subfigure{\includegraphics[width=.33\textwidth]{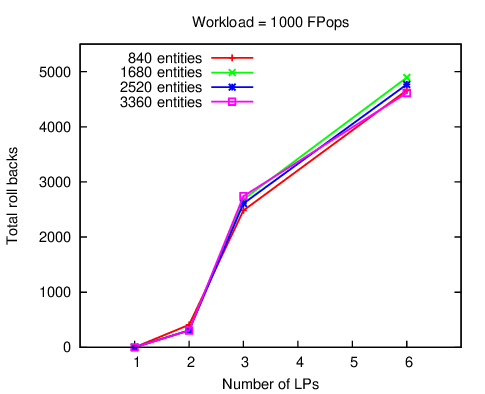}}
\subfigure{\includegraphics[width=.33\textwidth]{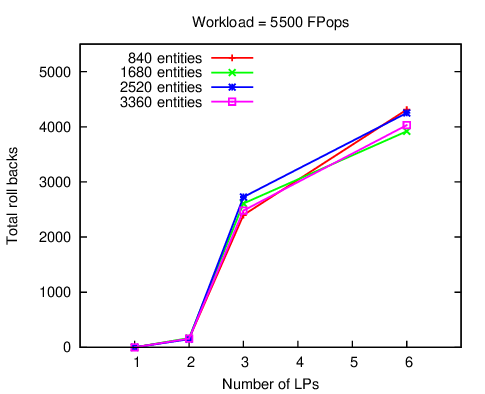}}
\subfigure{\includegraphics[width=.33\textwidth]{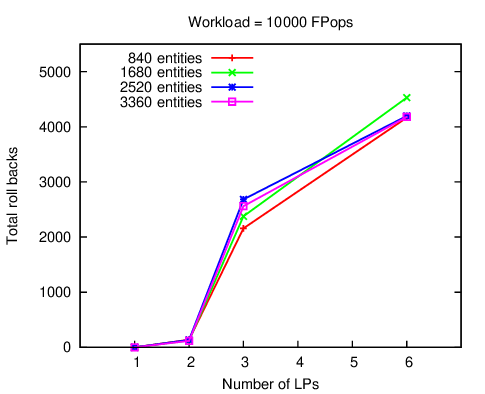}}
\caption{Total number of rollbacks on the distributed memory cluster
  as a function of the number of LPs (lower is better); the~\ac{GVT}
  is computed every 1$s$ of wall clock
  time}\label{fig:rollbacks_distributed_1s}
\end{figure*}

\paragraph{ErlangTW on Distributed Memory}
The distributed memory system is the research cluster of the~\ac{PADS}
group at the University of Bologna. We used three machines,
\texttt{cassandra}, \texttt{cerbero} and \texttt{chernobog} whose
configuration is shown in Table~\ref{tab:testbed}. We performed
experiments with $L=1, 2, 3, 6$. For $L=1$, the~\ac{LP} executed on
\texttt{cassandra}; for $L=2$, one~\ac{LP} executed on
\texttt{cassandra} and the other one on \texttt{cerbero}.  For $L=3$
we run a single~\ac{LP} on each of the three machines. Finally, when
$L=6$ we executed two~\acp{LP} on each of the three machines. 

Figure~\ref{fig:speedup_distributed_5s} shows the speedup of the PHOLD
model, measured on our distributed memory cluster. Thanks to the
Erlang language, it was possible to execute the exact same
implementation which was tested on the shared memory machine. Again,
each value is obtained by averaging 30 simulation runs. The most
prominent feature of these figures is the superlinear speedup which
occurs with $L=2$ and $L=3$ \acp{LP}. As in most of these situations,
this superlinear speedup can be explained by the fact that the machine
used for the test with $L=1$ (\texttt{cassandra}) has limited memory,
and therefore makes use of virtual memory during the simulation. To
confirm this hypothesis, we reduced the amount of memory required by
the PHOLD model by reducing the wall clock time between~\ac{GVT}
calculations. Recall from Section~\ref{sec:tw} that, once the~\ac{GVT}
is known, each~\ac{LP} can discard logs for events executed before
the~\ac{GVT}, since these events will be never rolled back. Therefore,
increasing the frequency of~\ac{GVT} calculation results in a reduced
memory footprint of the simulation model, at the cost of a higher
number of communications. The test shown in
Figure~\ref{fig:speedup_distributed_5s} were done with the~\ac{GVT}
computed every 5$s$ of wall clock time; reducing this interval to 1$s$
produces the more reasonable results shown in
Figure~\ref{fig:speedup_distributed_1s}. 

Scalability on the distributed memory cluster is quite poor, as
confirmed by the efficiency shown in
Figure~\ref{fig:efficiency_distributed_1s}. This result can be
explained by observing that PADS applications often exhibit low
computation / communication ratio, and in our distributed memory
testbed the communication network uses the standard Gigabit Ethernet
protocol which suffers from non negligible latency. Note from
Figure~\ref{fig:efficiency_distributed_1s} that scalability and
efficiency are particularly poor for low workload intensities (1000
and 5500 FPops) and for low number of entities. In these situations
PHOLD is communication bound, and the latency introduced by the
commodity LAN severely impacts on the overall performance.

Since our cluster includes heterogeneous machines, there load is not
evenly balanced across the~\acp{LP}, and this generates a large number
of antimessages. In Figure~\ref{fig:rollbacks_distributed_1s} we plot
the mean total number of rollbacks as a function of the number
of~\acp{LP} $L$. The number of rollbacks sharply increases from $L=2$
to $L=3$, and this can be explained by the fact that for
\verb+cassandra+ and \verb+cerbero+ have a similar hardware configuration,
while \verb+chernobog+ (which is used when $L=3$ and $L=6$) is much
more powerful. As in the shared memory case, the faster~\acp{LP} is
prone to produce a large number of stragglers which generate a cascade
of rollbacks.

\section{Conclusion and future work}\label{sec:conclusions}

In this paper we described ErlangTW, an implementation of the Time
Warp protocol for parallel and distributed simulations implemented in
Erlang. ErlangTW allows the same simulation model to be executed
(unmodified) either on single-core, multicore and distributed
computing architectures. We described the prototype implementation of
ErlangTW, and analyzed its scalability on a multicore, shared memory
machine and on a small distributed memory cluster using the PHOLD
benchmark.

Results show that Erlang provides a good framework to build
simulators, thanks to its powerful language features and virtual
machine facilities; furthermore, Erlang's transparent message
brokering system greatly simplifies the development of complex
distributed applications, such as~\ac{PADS}. Performance of the PHOLD
benchmark show that scalability and efficiency on shared memory
architectures are very good, while distributed memory architectures
are less friendly--performance wise--to these kinds of applications.

As seen before, the communication overhead of the distributed
execution environment has a big impact on the simulator performances
and the Time Warp synchronization algorithm reacts badly to imbalances
in the execution architecture (e.g.~CPUs with very different speeds or
presence of background load). Both these problems can be addressed
using nice features provided by Erlang: the serialization of objects
and data structures, and code migration. Thanks to this, it is
possible to implement the transfer of simulated entities across
different~\acp{LP} or even moving a whole~\ac{LP} on a different CPU,
all at runtime. In this way, the ErlangTW simulator would be able to
reduce the communication cost by adaptively clustering highly
interacting entities within the same~\ac{LP}. Furthermore, it will be
possible to implement other advanced forms of
load-balancing~\cite{gda-ijspm-2009} to speed up the execution and to
reduce the number of roll-backs. This will permit the implementation
of new adaptive simulators that can change their configuration at
runtime. To further enhance the performance of ErlangTW, we will
exploit additional parallelization of the~\ac{LP}, by decoupling
message dispatching from entity management using separate~LWT.

\section*{Source Code Availability}\label{sec:source}
The ErlangTW Simulator is released under the GNU General Public License (GPL) 
version 2 and can be freely downloaded from
\url{http://pads.cs.unibo.it/}

\bibliographystyle{abbrvnat}
\bibliography{Erlang-Warp}

\begin{thebibliography}{34}
\providecommand{\natexlab}[1]{#1}
\providecommand{\url}[1]{\texttt{#1}}
\expandafter\ifx\csname urlstyle\endcsname\relax
  \providecommand{\doi}[1]{doi: #1}\else
  \providecommand{\doi}{doi: \begingroup \urlstyle{rm}\Url}\fi

\bibitem[sim(2012)]{sim-diasca}
{Sim-Diasca}.
\newblock \url{http://www.sim-diasca.org/}, 2012.

\bibitem[Andersson(1999)]{And99}
A.~Andersson.
\newblock General balanced trees.
\newblock \emph{J. Algorithms}, 30\penalty0 (1):\penalty0 1--18, Jan. 1999.
\newblock ISSN 0196-6774.
\newblock \doi{10.1006/jagm.1998.0967}.

\bibitem[Armstrong(2007)]{Armstrong07}
J.~Armstrong.
\newblock \emph{Programming Erlang: Software for a Concurrent World}.
\newblock Pragmatic Bookshelf, 2007.
\newblock ISBN 193435600X, 9781934356005.

\bibitem[Bagrodia et~al.(1998)Bagrodia, Meyer, Takai, Chen, Zeng, Martin, and
  Song]{Bagrodia1998}
R.~Bagrodia, R.~Meyer, M.~Takai, Y.-A. Chen, X.~Zeng, J.~Martin, and H.~Y.
  Song.
\newblock Parsec: a parallel simulation environment for complex systems.
\newblock \emph{Computer}, 31\penalty0 (10):\penalty0 77 --85, oct 1998.
\newblock ISSN 0018-9162.
\newblock \doi{10.1109/2.722293}.

\bibitem[Bononi et~al.(2006)Bononi, Bracuto, D'Angelo, and
  Donatiello]{gda-dsrt-2006}
L.~Bononi, M.~Bracuto, G.~D'Angelo, and L.~Donatiello.
\newblock Exploring the effects of {Hyper-Threading} on parallel simulation.
\newblock In \emph{Proceedings of the 10th IEEE international symposium on
  Distributed Simulation and Real-Time Applications}, pages 257--260,
  Washington, DC, USA, 2006. IEEE Computer Society.
\newblock ISBN 0-7695-2697-7.
\newblock \doi{10.1109/DS-RT.2006.18}.

\bibitem[Carlson and Tronje(1995)]{sim94}
B.~Carlson and S.~Tronje.
\newblock Sim94--a concurrent simulator for plan-driven troops.
\newblock Technical report, Uppsala Universitet, Sweden, Feb. 15 1995.

\bibitem[Carothers and Fujimoto(2000)]{Carothers2000}
C.~D. Carothers and R.~M. Fujimoto.
\newblock Efficient execution of time warp programs on heterogeneous, {NOW}
  platforms.
\newblock \emph{IEEE Trans. Parallel Distrib. Syst.}, 11\penalty0 (3):\penalty0
  299--317, Mar. 2000.
\newblock ISSN 1045-9219.
\newblock \doi{10.1109/71.841745}.

\bibitem[Chen and Szymanski(2005)]{chen2005}
G.~Chen and B.~Szymanski.
\newblock {DSIM}: scaling time warp to 1,033 processors.
\newblock In \emph{Simulation Conference, 2005 Proceedings of the Winter}, page
  10 pp., dec. 2005.
\newblock \doi{10.1109/WSC.2005.1574269}.

\bibitem[Dahl and Nygaard(1966)]{Dahl1966}
O.-J. Dahl and K.~Nygaard.
\newblock {SIMULA}: an {ALGOL}-based simulation language.
\newblock \emph{Commun. ACM}, 9\penalty0 (9):\penalty0 671--678, Sept. 1966.
\newblock ISSN 0001-0782.
\newblock \doi{10.1145/365813.365819}.

\bibitem[D'Angelo(2011)]{Gda11}
G.~D'Angelo.
\newblock Parallel and distributed simulation from many cores to the public
  cloud.
\newblock In \emph{High Performance Computing and Simulation (HPCS), 2011
  International Conference on}, pages 14--23, July 2011.
\newblock \doi{10.1109/HPCSim.2011.5999802}.

\bibitem[D'Angelo and Bracuto(2009)]{gda-ijspm-2009}
G.~D'Angelo and M.~Bracuto.
\newblock Distributed simulation of large-scale and detailed models.
\newblock \emph{International Journal of Simulation and Process Modelling
  (IJSPM)}, 5\penalty0 (2):\penalty0 120--131, 2009.
\newblock ISSN 1740-2123.

\bibitem[D'Angelo et~al.(2012)D'Angelo, Ferretti, and Marzolla]{gda12}
G.~D'Angelo, S.~Ferretti, and M.~Marzolla.
\newblock Time warp on the {Go}.
\newblock In \emph{Proc. Simutools 2012 - Fifth International Conference on
  Simulation Tools and Techniques}, pages 249--255, Desenzano, Italy, Mar.19
  2012.

\bibitem[Fujimoto(1989)]{Fujimoto:1989:PDE:76738.76741}
R.~M. Fujimoto.
\newblock Parallel discrete event simulation.
\newblock In \emph{Proceedings of the 21st conference on Winter simulation},
  WSC '89, pages 19--28, New York, NY, USA, 1989. ACM.
\newblock ISBN 0-911801-58-8.

\bibitem[Fujimoto(1990)]{Fuj90}
R.~M. Fujimoto.
\newblock Performance of time warp under synthetic workloads.
\newblock In \emph{Proc. SCS Multiconference on Distributed Simulation}, pages
  23–--28, 1990.

\bibitem[Fujimoto(2000)]{Fuj00}
R.~M. Fujimoto.
\newblock \emph{Parallel and distributed simulation systems}.
\newblock Wiley series on parallel and distributed computing. Wiley, 2000.
\newblock ISBN 9780471183839.

\bibitem[Gordon(1978)]{Gordon1978}
G.~Gordon.
\newblock The development of the general purpose simulation system (gpss).
\newblock \emph{SIGPLAN Not.}, 13\penalty0 (8):\penalty0 183--198, Aug. 1978.
\newblock ISSN 0362-1340.
\newblock \doi{10.1145/960118.808382}.

\bibitem[Jefferson(1985)]{Jef85}
D.~R. Jefferson.
\newblock Virtual time.
\newblock \emph{ACM Trans. Program. Lang. Syst.}, 7\penalty0 (3):\penalty0
  404--425, July 1985.
\newblock ISSN 0164-0925.
\newblock \doi{10.1145/3916.3988}.

\bibitem[Jones(1986)]{Jon86}
D.~W. Jones.
\newblock An empirical comparison of priority-queue and event-set
  implementations.
\newblock \emph{Commun. ACM}, 29\penalty0 (4):\penalty0 300--311, Apr. 1986.
\newblock ISSN 0001-0782.
\newblock \doi{10.1145/5684.5686}.

\bibitem[Lamport(1978)]{Lamport78}
L.~Lamport.
\newblock Time, clocks, and the ordering of events in a distributed system.
\newblock \emph{Commun. ACM}, 21\penalty0 (7):\penalty0 558--565, July 1978.
\newblock ISSN 0001-0782.
\newblock \doi{10.1145/359545.359563}.

\bibitem[Low et~al.(1999)Low, Lim, Cai, Huang, Hsu, Jain, and Turner]{Low99}
Y.-H. Low, C.-C. Lim, W.~Cai, S.-Y. Huang, W.-J. Hsu, S.~Jain, and S.~J.
  Turner.
\newblock Survey of languages and runtime libraries for parallel discrete-event
  simulation.
\newblock \emph{SIMULATION}, 72\penalty0 (3):\penalty0 170--186, 1999.
\newblock \doi{10.1177/003754979907200309}.

\bibitem[Marr et~al.(2002)Marr, Binns, Hill, Hinton, Koufaty, Miller, and
  Upton]{HT}
D.~T. Marr, F.~Binns, D.~L. Hill, G.~Hinton, D.~A. Koufaty, A.~J. Miller, and
  M.~Upton.
\newblock {Hyper-Threading Technology Architecture and Microarchitecture}.
\newblock \emph{Intel Technology Journal}, 6\penalty0 (1), Feb. 2002.

\bibitem[Misra(1986)]{misra86}
J.~Misra.
\newblock Distributed discrete event simulation.
\newblock \emph{ACM Computing Surveys}, 18\penalty0 (1):\penalty0 39--65, 1986.

\bibitem[Park and Miller(1988)]{PM88}
S.~K. Park and K.~W. Miller.
\newblock Random number generators: good ones are hard to find.
\newblock \emph{Commun. ACM}, 31\penalty0 (10):\penalty0 1192--1201, Oct. 1988.
\newblock ISSN 0001-0782.
\newblock \doi{10.1145/63039.63042}.

\bibitem[Perumalla(2005)]{Perumalla:2005:MPS:1069810.1070161}
K.~S. Perumalla.
\newblock $\mu$sik - a micro-kernel for parallel/distributed simulation
  systems.
\newblock In \emph{Proceedings of the 19th Workshop on Principles of Advanced
  and Distributed Simulation}, PADS '05, pages 59--68, Washington, DC, USA,
  2005. IEEE Computer Society.
\newblock ISBN 0-7695-2383-8.

\bibitem[Perumalla(2006)]{perumalla2007}
K.~S. Perumalla.
\newblock Parallel and distributed simulation: traditional techniques and
  recent advances.
\newblock In L.~F. Perrone, B.~Lawson, J.~Liu, and F.~P. Wieland, editors,
  \emph{Proceedings of the Winter Simulation Conference WSC 2006, Monterey,
  California, USA, December 3-6, 2006}, pages 84--95. WSC, 2006.
\newblock ISBN 1-4244-0501-7.
\newblock \doi{10.1145/1218112.1218132}.

\bibitem[Perumalla et~al.(2011)Perumalla, Park, and Tipparaju]{Perumalla11}
K.~S. Perumalla, A.~J. Park, and V.~Tipparaju.
\newblock {GVT} algorithms and discrete event dynamics on 129k+ processor
  cores.
\newblock In \emph{18th International Conference on High Performance Computing,
  HiPC 2011, Bengaluru, India, December 18-21, 2011}, pages 1--11. IEEE, 2011.
\newblock ISBN 978-1-4577-1951-6.
\newblock \doi{10.1109/HiPC.2011.6152725}.

\bibitem[PRIME()]{prime}
PRIME.
\newblock {Parallel Real-time Immersive network Modeling Environment - PRIME}.
\newblock \url{https://www.primessf.net}, 2011.

\bibitem[Rice et~al.(2005)Rice, Markowitz, Marjanski, and Bailey]{Rice2005}
S.~V. Rice, H.~M. Markowitz, A.~Marjanski, and S.~M. Bailey.
\newblock The {SIMSCRIPT III} programming language for modular object-oriented
  simulation.
\newblock In \emph{Proceedings of the 37th conference on Winter simulation},
  WSC '05, pages 621--630. Winter Simulation Conference, 2005.
\newblock ISBN 0-7803-9519-0.

\bibitem[Richardson and Pugh(1981)]{Richardson1981}
G.~P. Richardson and A.~L. Pugh.
\newblock \emph{Introduction to System Dynamics Modeling with Dynamo}.
\newblock MIT Press, Cambridge, MA, USA, 1981.
\newblock ISBN 0262181029.

\bibitem[Samadi et~al.(1987)Samadi, Muntz, and Parker]{SMP87}
B.~Samadi, R.~Muntz, and D.~Parker.
\newblock A distributed algorithm to detect a global state of a distributed
  simulation system.
\newblock In \emph{Proc. IFIP Conference on Distributed Processing}.
  North-Holland, 1987.

\bibitem[Shaw et~al.(2008)Shaw, Deneroff, Dror, Kuskin, Larson, Salmon, Young,
  Batson, Bowers, Chao, Eastwood, Gagliardo, Grossman, Ho, Ierardi,
  Kolossv\'{a}ry, Klepeis, Layman, McLeavey, Moraes, Mueller, Priest, Shan,
  Spengler, Theobald, Towles, and Wang]{Shaw08}
D.~E. Shaw, M.~M. Deneroff, R.~O. Dror, J.~S. Kuskin, R.~H. Larson, J.~K.
  Salmon, C.~Young, B.~Batson, K.~J. Bowers, J.~C. Chao, M.~P. Eastwood,
  J.~Gagliardo, J.~P. Grossman, C.~R. Ho, D.~J. Ierardi, I.~Kolossv\'{a}ry,
  J.~L. Klepeis, T.~Layman, C.~McLeavey, M.~A. Moraes, R.~Mueller, E.~C.
  Priest, Y.~Shan, J.~Spengler, M.~Theobald, B.~Towles, and S.~C. Wang.
\newblock Anton, a special-purpose machine for molecular dynamics simulation.
\newblock \emph{Commun. ACM}, 51\penalty0 (7):\penalty0 91--97, July 2008.
\newblock ISSN 0001-0782.
\newblock \doi{10.1145/1364782.1364802}.

\bibitem[Steinman and Wong(2003)]{Steinman:2003:SPF:824475.825880}
J.~S. Steinman and J.~W. Wong.
\newblock The {SPEEDES} persistence framework and the standard simulation
  architecture.
\newblock In \emph{Proceedings of the seventeenth workshop on Parallel and
  distributed simulation}, PADS '03, pages 11--, Washington, DC, USA, 2003.
  IEEE Computer Society.
\newblock ISBN 0-7695-1970-9.

\bibitem[Steinman et~al.(2008)Steinman, Lammers, Valinski, Roth, and
  Words]{Steinman_08s-siw-025warpiv}
J.~S. Steinman, C.~N. Lammers, M.~E. Valinski, K.~Roth, and K.~Words.
\newblock Simulating parallel overlapping universes in the fifth dimension with
  {HyperWarpSpeed} implemented in the {WarpIV} kernel.
\newblock In \emph{Proceedings of the Simulation Interoperability Workshop},
  SIW '08, 2008.

\bibitem[Tay et~al.(2003)Tay, Tan, and Shenoy]{1261535}
S.~Tay, G.~Tan, and K.~Shenoy.
\newblock Piggy-backed time-stepped simulation with 'super-stepping'.
\newblock In \emph{Simulation Conference, 2003. Proceedings of the 2003
  Winter}, volume~2, pages 1077 -- 1085 vol.2, dec. 2003.

\end{thebibliography}

\end{document}